\documentclass[a4paper,11pt]{article}
\usepackage{pos}
\usepackage{pos}
\usepackage{multirow}
\usepackage{upgreek}
\usepackage{subcaption}
\usepackage{wrapfig}
\usepackage{enumitem}
\setenumerate{itemsep=0mm}
\usepackage{verbatim}

\title{Status and Future Prospects of the KASCADE Cosmic-ray Data Centre KCDC}
 \ShortTitle{KCDC}

\author*[a]{Andreas Haungs}
\author[a]{Donghwa Kang} 
\author[a]{Katrin Link} 
\author[a]{Frank Polgart}
\author[a]{Victoria Tokareva} 
\author[a]{Doris Wochele} 
\author[a]{Jürgen Wochele}

\affiliation[a]{Karlsruhe Institute of Technology, Institute for Astroparticle Physics, 76021 Karlsruhe, Germany}

\emailAdd{andreas.haungs@kit.edu}

\abstract{
KCDC, the 'KASCADE Cosmic-ray Data Centre', is a web-based interface where initially the scientific data from the completed air-shower experiment KASCADE-Grande was made available for the astroparticle community as well as for the interested public. Over the past 7 years, we have continuously extended the data shop with various releases and increased both the number of detector components from the KASCADE-Grande experiment and the data sets and corresponding simulations. With the latest releases we added a new and independent data shop for a specific KASCADE-Grande event selection and by that created the technology for integrating further data shops and data of other experiments, like the data of the air-shower experiment MAKET-ANI in Armenia. In addition, we made available educational examples how to use the data, more than 100 cosmic ray energy spectra from various experiments, and recently attached a public server with access to Jupyter notebooks. In this paper we present a brief history of KCDC, the main features of the recent release as well as will discuss future development plans.
}

\FullConference{37$^{\rm{th}}$ International Cosmic Ray Conference (ICRC 2021)\\
		July 12th -- 23rd, 2021\\
		Online -- Berlin, Germany}

\begin{document}
\maketitle

\section{Introduction}

\begin{wrapfigure}{l}{0.45\textwidth}
  \begin{center}
    \includegraphics[width=0.43\textwidth]{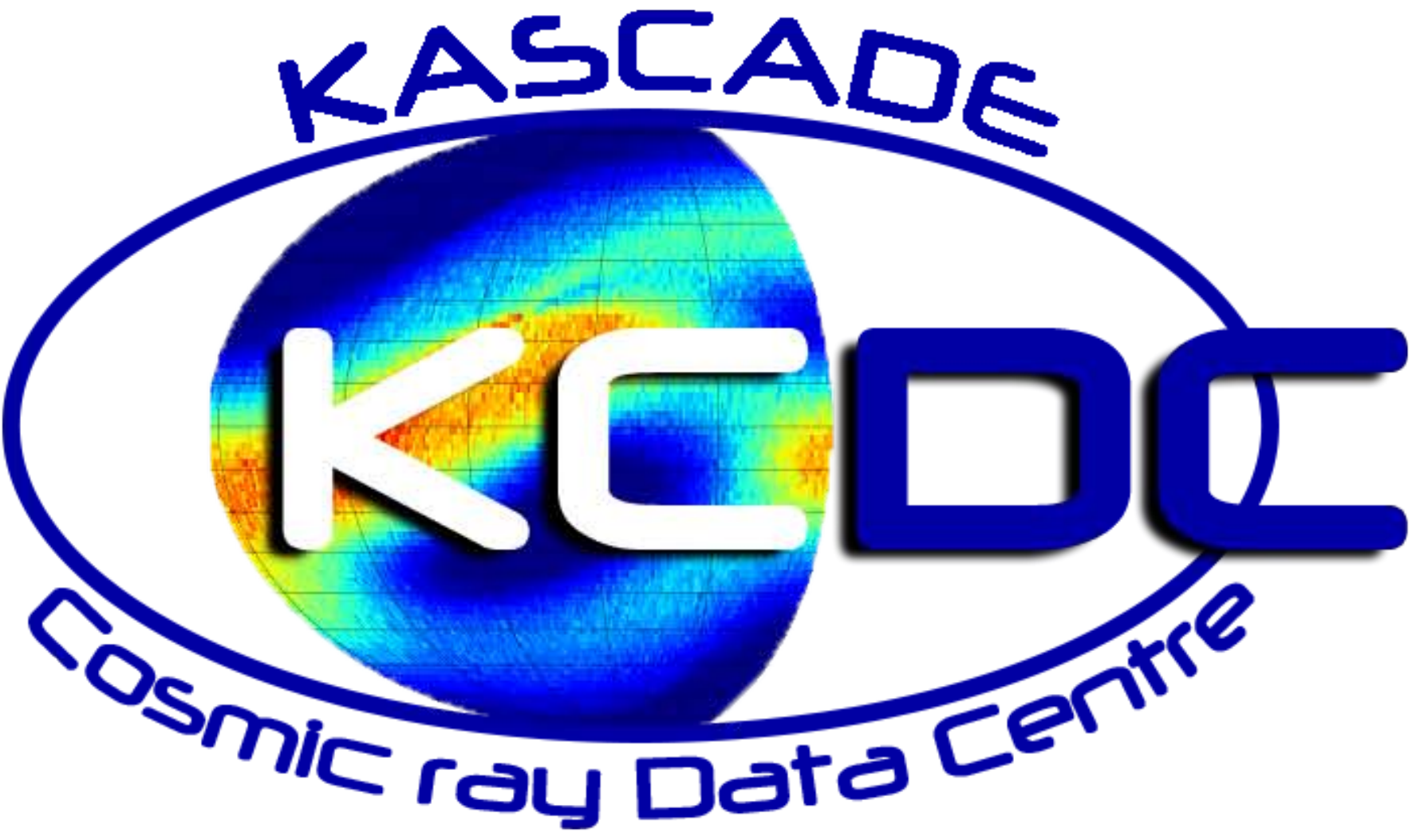}
  \end{center}
 \caption{Logo of KCDC. The link to KCDC is: \large{\bf \url{https://kcdc.iap.kit.edu}}.}  
 \label{fig_logo}
\end{wrapfigure}
KCDC, the 'KASCADE Cosmic-ray Data Centre' \url{https://kcdc.iap.kit.edu/}~\cite{kcdc_paper}, is a web-based interface where air-shower data is made publicly available. Over the past 7 years, we have continuously extended the platform. 
The aim of the project KCDC is the installation and establishment of a public data centre for high-energy astroparticle physics based on the data of the KASCADE experiment (logo of KCDC, see~Figure~\ref{fig_logo}). 
KASCADE and KASCADE-Grande~\cite{kcdc_paper,kang_kascade_icrc2021} -- as initial data provider -- was a quite successful large detector array for measuring high-energy cosmic rays 
via the detection of extensive air showers (EAS).  KASCADE recorded data during more than 20 years on 
site of the KIT, Campus North, Karlsruhe, Germany (formerly Forschungszentrum Karlsruhe) at $49.1^\circ$N, $8.4^\circ$E, and $110\,$m a.s.l. 

\section{Short History of Past Releases}

With the establishment of a public data centre for high-energy astroparticle physics, we entered a new territory, where no self-contained concepts were available, how the data should be treated and prepared that they could be used reasonably outside the KASCADE collaboration. From the first release, we tried to meet three basic requirements:

\begin{itemize}[itemsep=0pt,topsep=0pt]
\item \textit{KCDC as data provider}: granting free and unlimited open access to KASCADE data,
\item \textit{KCDC as information platform}: providing a detailed experiment description and sufficient meta information on the data and the analysis procedures,
\item \textit{KCDC as long-term digital data archive}: KCDC serves not only as a software and data archive for the collaboration, but also for the public.
\end{itemize}

With the first release in November 2013 (called \textbf{WOLF359}), we published 158 Mio events with seven reconstructed parameters called quantities, based on data analyses of the original KASCADE detector array  only. With every major release, we added more data sets and/or more KASCADE-Grande detector components. 

With the release \textbf{VULCAN} in November 2014, we switched from SQL to NoSQL data base (MongoDB) and added two more KASCADE quantities from the KASCADE Hadron Calorimeter.

With \textbf{MERIDIAN}, released in November 2015, the  measured detector data from each of the 252 KASCADE Array detector stations were published for the first time, the ‘Energy Densities’ and the ‘Arrival Times’ per station, which enlarged the database by a factor of 100. 
In our software package KAOS, which is the basic framework for KCDC, two new plugins handling the KCDC ‘Publications’ and first ‘Spectra’ data of related cosmic-ray experiments for download were included.

A big change has been introduced with the release of \textbf{NABOO} in February 2017. In addition to the newly added GRANDE detector component, all data from KASCADE and GRANDE, recorded between 1997 and 2013 were published increasing the number of events to 433 million with roughly 800 data words per event. The database thus reached a size of 3TB, which made it virtually unavoidable to completely rewrite the back-end programming to handle this amount of data. Furthermore, the matching CORSIKA simulations for KASCADE and GRANDE for six different high-energy hadronic interaction models were released. The simulation data were provided in the same data structure as the measured data. The data arrays of ‘Energy Density’ per station was replaced by the ‘Energy Deposits’ which is closer to the measured data and offers more analysis options for the users. In addition, the sample of available spectra at the KCDC web portal has been extended to 88 published spectra from 21 different cosmic-ray experiments.

\begin{figure}[t]
\centering
\includegraphics[width=0.55\textwidth]{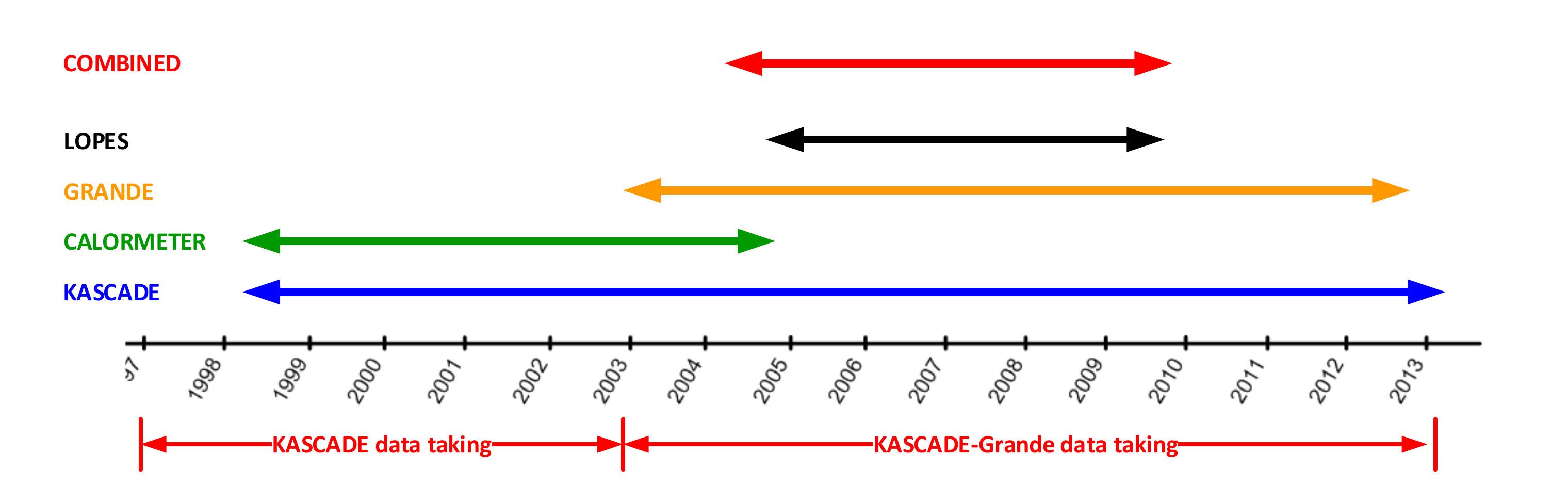}
\includegraphics[width=0.40\textwidth]{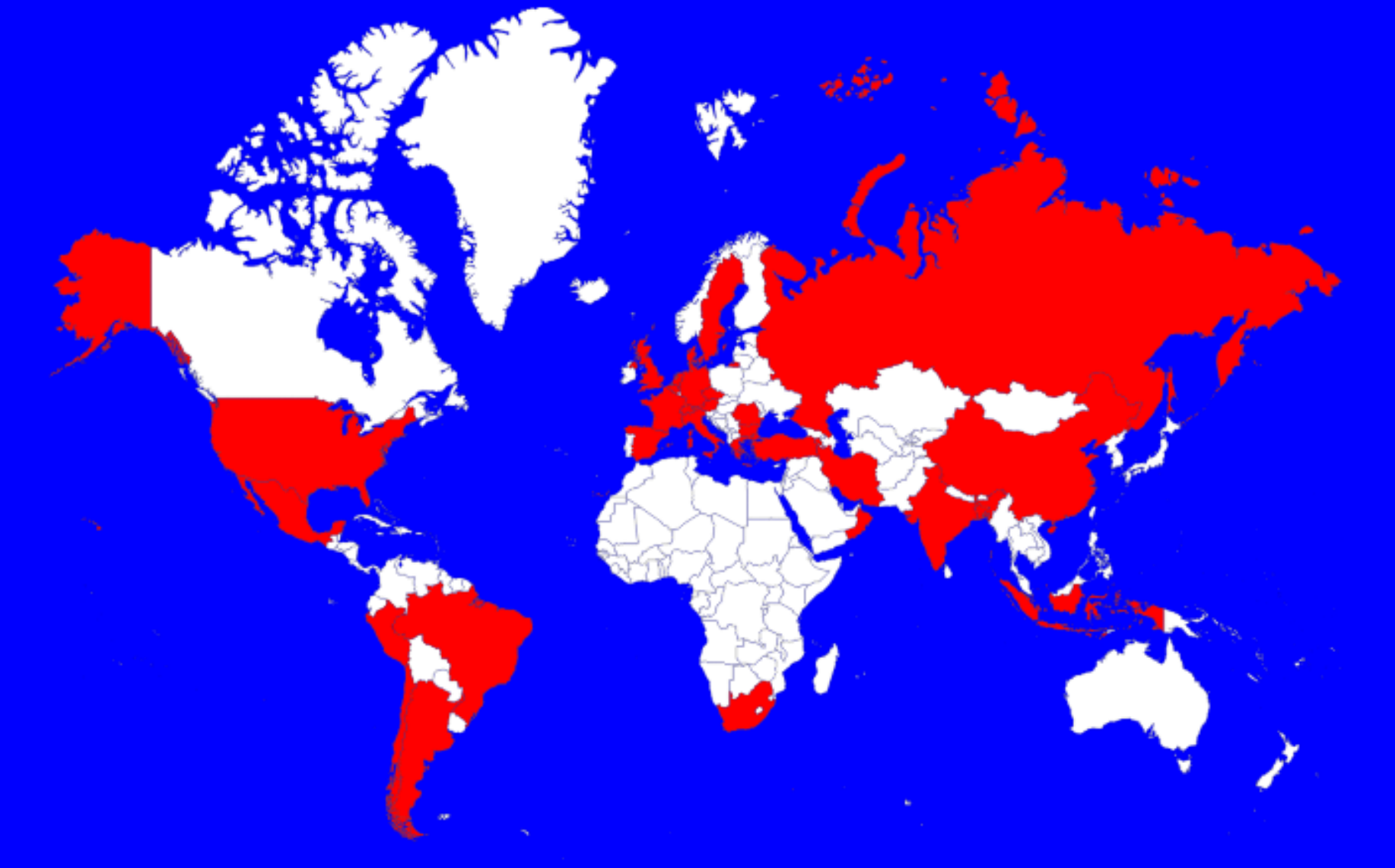}
\caption{Left: Timeline of ‘active times’ of the KASCADE-Grande detector components as published at KCDC. Right: World map showing the distribution of KCDC users (red).} 
\label{timeline}
\end{figure}

With the release \textbf{OCEANUS} in November 2019 we moved the DataShop Mongo DB to a sharded cluster speeding up the processing time for user requests by roughly a factor of 50. Furthermore, the data of the Radio LOPES detectors~\cite{LOPES,lopes-new} were added as a new detector component within the KASADE DataShop. Fig.~\ref{timeline} shows on a timeline the data acquisition times of the data sets published in KCDC for the four detector components and the COMBINED data analyses.

In between these ‘major releases’ we published some ‘minor releases’ when, for example, new ‘Spectra’ have been added or when bug-fixes were necessary.

From the very beginning, a large interest from the community was given, proved, e.g.~by our anonymous monitoring of the access to the portal. Up to now more than 330 users registered from more than 30 countries distributed over five continents (see fig.~\ref{timeline}). We track page views and downloads with MATOMO analytics~\cite{MATOMO} to learn about the customers needs. Furthermore, we have to report usage statistics anonymously to the public financiers. As communication platform serves an Email list for the KCDC subscribers as well as social networks, like Twitter \url{https://twitter.com/KCDC_KIT}.

\section{Upgrade to Second DataShop}

Our declared goal from the very beginning was to implement also data from other cosmic-ray experiments into KCDC as new DataShops. On the one hand, these shops should take into account the specific properties of the respective experiments and, on the other hand, they should be able to be integrated into the KCDC framework. 

With the release \textbf{PENTARUS} in May 2020 we added a second DataShop called \textbf{‘COMBINED’}, where the data from the joint analysis of the KASCADE and the GRANDE detector arrays were made publicly available. Thus ’COMBINED’ is a subsample of the KASCADE data, analysed in a completely different way, which made it necessary to set up a new data base. The combined detector output has the quality of a stand-alone experiment, not of an additional component. Data taken by KASCADE and its extension GRANDE were analysed more or less independently of each other before the arrays were combined on raw data level. The aim of the combined analysis was to utilise an improved reconstruction to get one single, consistent spectrum in the energy range of $10^{15}\,$eV to $10^{18}\,$eV. The focus is on the mass composition, which is one of the most important sources of information needed to restrict astrophysical models on the origin and propagation of cosmic rays. With the improved reconstruction of the extensive air showers, a study of the elemental composition of high-energy cosmic rays is possible in a more detailed way.

Adding ‘COMBINED’ as a new DataShop to KCDC had many implications for the existing web portal:
\begin{itemize}[itemsep=0pt,topsep=0pt]
\item the back-end programming was extended to host more DataShops,
\item a new MongoDB was set up and filled with the ‘COMBINED’ data sets,
\item matching CORSIKA simulations were generated for ‘COMBINED’,
\item new ‘Preselections’ were added,
\item new documentations for ‘COMBINED’ and ‘COMBINED-Simulations’ were  provided,
\item most of the static and dynamic web pages needed refurbishing,
\item a new menu item called ‘Materials’ was added to host the manuals and the programming tools for all DataShops,
\item the ‘Simulations’ and ‘Preselections’ pages were completely redesigned.
\end{itemize}

With these changes we have created the basis to incorporate other cosmic-ray experiments into KCDC in future without having to make major changes to the backend programming, especially since the definition of the new detector components and the quantities is done in a web-based ‘Admin Interface’.

\section{Maket-Ani DataShop}
In the latest release \textbf{SKARAGAN} published February 2021, we introduced for the first time a DataShop not related to the KASCADE-Grande experiment, the \textbf{‘Maket-Ani'} DataShop. 
\textbf{Maket-Ani}~\cite{apj_603-2004,ap_28-2007} is an extensive air shower experiment array to study the cosmic ray primary flux in the ‘knee’ region of the primary cosmic ray spectrum.The Maket-Ani detector is placed on Mt.Aragats (Aragats  Cosmic  Ray  Observatory, Armenia),  3200\,m  above sea  level at 40$^{\circ}$25’N, 44$^{\circ}$15’E operating at an atmospheric depth of $\approx 700\,$g/cm$^{2}$. At this altitude the shape of  the  showers  is  not  distorted  by  the  attenuation  in  the  terrestrial  atmosphere  and it  is possible to reliably reconstruct EAS size and shape.
\begin{figure}[t]
\centering
\includegraphics[width=0.45\textwidth]{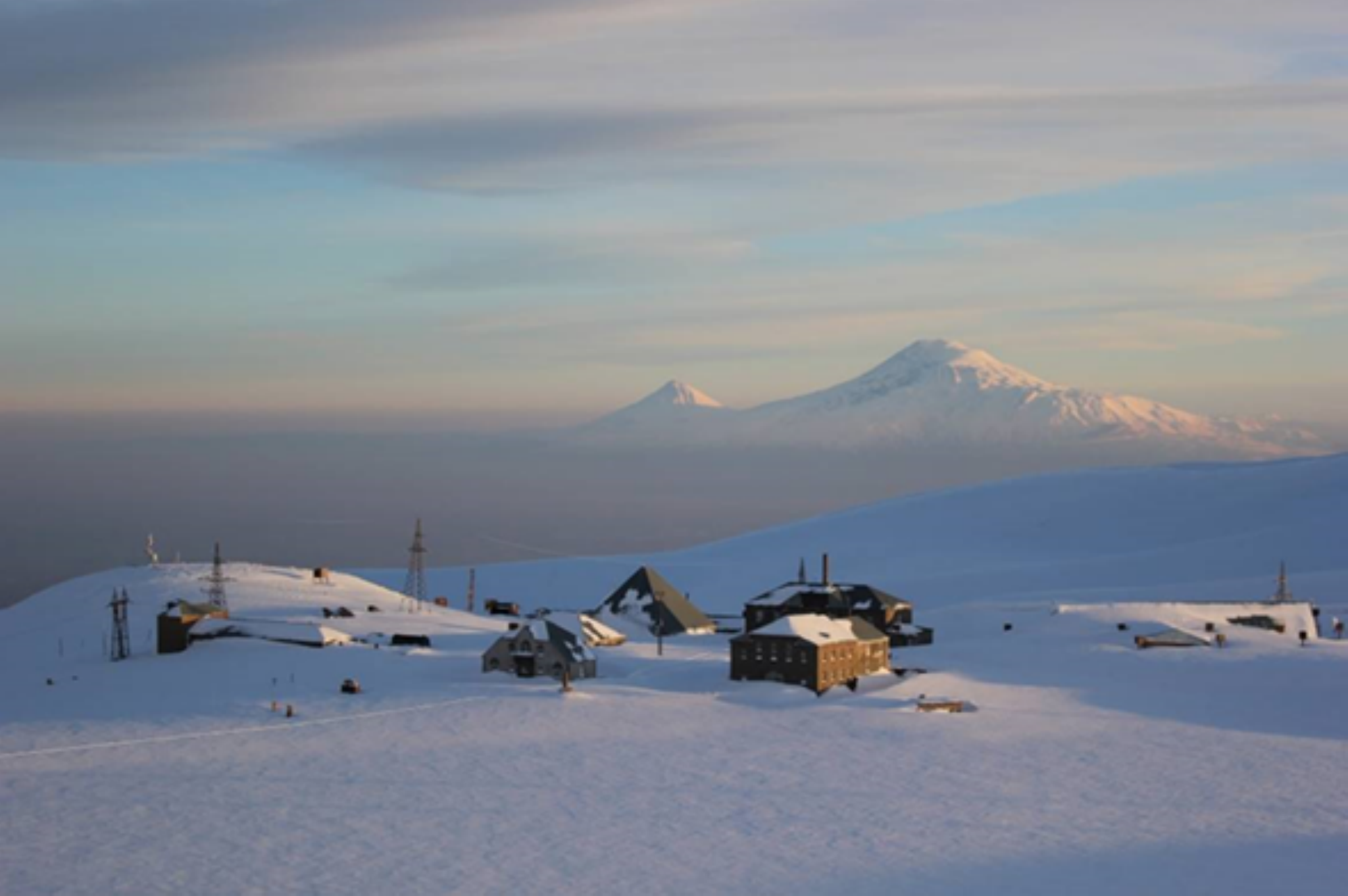}
\includegraphics[width=0.45\textwidth]{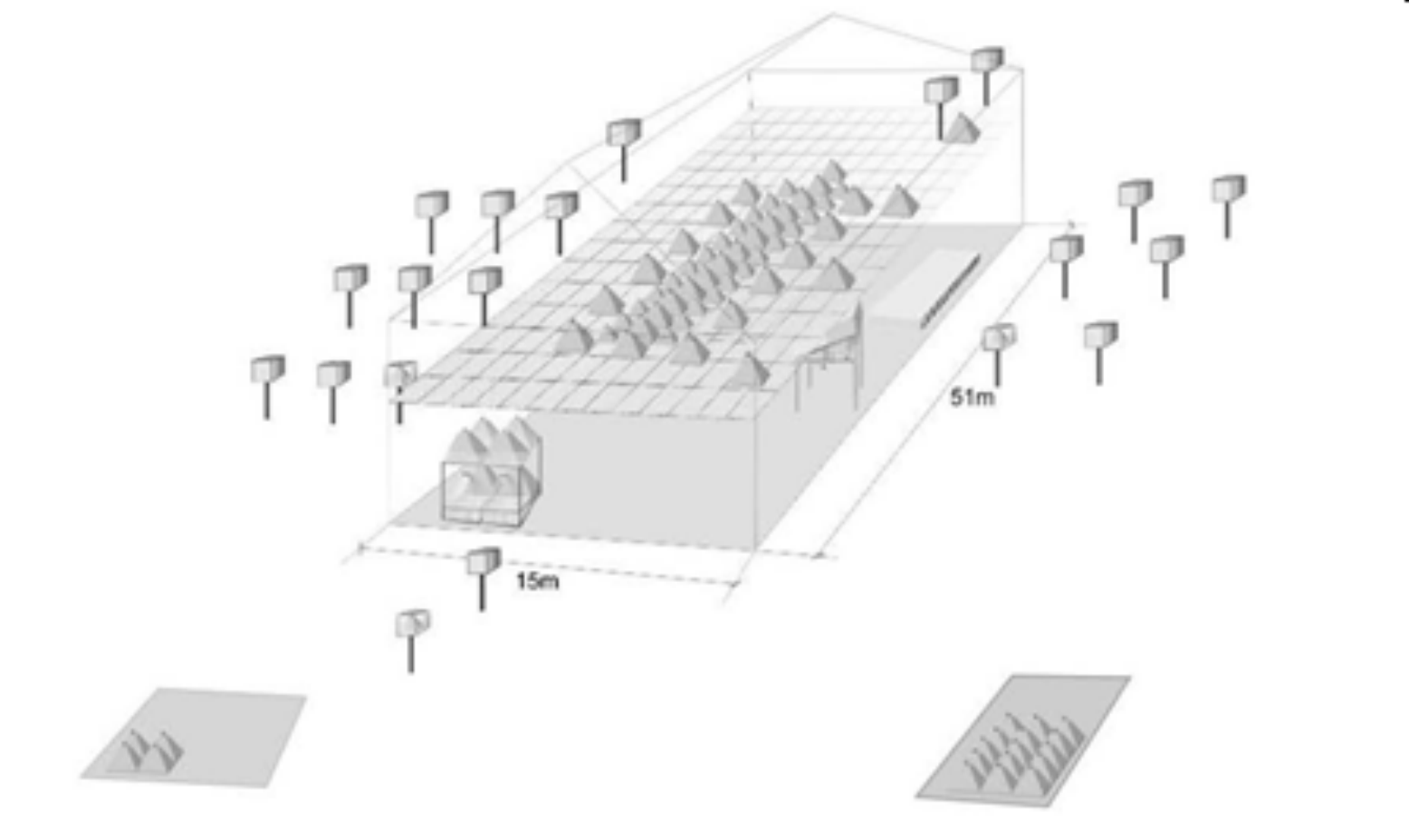}
\caption{Left: View of the Maket-Ani location at Mount Aragats. Right: Maket-Ani detector setup.} 
\label{MA_location}
\end{figure} 

The Maket-Ani surface array (fig.~\ref{MA_location}) consists of 92 particle detectors formed by plastic scintillators with a thickness of 5 cm. Twenty-four of them have an area of 0.09\,m$^{2}$ and 68 an area of 1\,m$^{2}$. The central part consists of 73 scintillation detectors, arranged in a rectangle of 85 x 65 m$^{2}$. Two peripheral points at a distance of 95 m and 65 m from the centre of the installation consist of 15 and 4 scintillators, respectively.

The average frequencies of the selected EAS are very stable and equal to $0.234 \pm 0.013\,$min$^{-1}$ for $Ne \geq 1.6 \cdot 10^{5}$; $0.056 \pm 0.006\,$min$^{-1}$ for $Ne \geq 4 \cdot 10^{5}$ and $0.013 \pm 0.003\,$min$^{-1}$ for $Ne \geq 10^{6}$.

During multiyear measurements, the detecting channels were continuously monitored. Data on background cosmic-ray spectra was collected for each detector. The slope of the spectra was used for detector calibration. The slope of background spectra is a very stable parameter and does not change even during very severe Forbush decreases, when the mean count rates can decrease as much as 20\%.

The Maket-Ani data in general are made publicly available via the ‘UNESCO Open Science recommendation’. The current data sets for KCDC, however, are taken from an ASCII file provided by the Maket-Ani collaboration directly. 
With the release \textbf{SKARAGAN} we published \textit{2.682.264} events recorded with the Maket-Ani detector array between 1.6.1997 and 22.3.2007 where quality cuts have been applied by the Maket-Ani collaboration. 
Besides the directly reconstructed data like core position (\textit{Xc, Yc}), shower direction (\textit{Ze, Az)}, shower age (\textit{Age}) and number of electrons (\textit{Ne}), some astronomical quantities are provided, like right ascension (\textit{Ra}) and declination (\textit{De}) as well as the local sidereal time at Maket-Ani location (\textit{St}).
From the recorded event time (\textit{date, time}) we generated a ‘Global Time’ (\textit{Gt}), which indicates the time in seconds elapsed since 1.1.1970. Additionally an UUID was added (\textit{UUID}) serving as a unique identifier for the respective event.
All of these events were put into a new MongoDB, most of which were provided with indices to enable quick data selection. The size of the data base is only about 1\,GB because no data arrays for the single detector stations are implemented.

\section{Jupyterhub}
A recent addition to our open data efforts is a public data analysis platform, based on the ubiquitous Jupyter notebook ecosystem, i.e. Jupyterhub plus Jupyterlab (\url{https://jupyter.iap.kit.edu}).
It is not uncommon for data sets generated by KCDC to be too large to be handled comfortably over slow internet connections, or limited end-user devices.
The reasonability to expect students or citizen-scientists to setup their own analysis framework is also debatable.
Our Jupyterlab supports analysis in python and C++, provides persistent storage and allows sharing among users or classes.
The notebooks have direct read access to the download area of KCDC, eliminating the necessity to copy huge chunks of data to a personal device, and potentially uploading it again to a third-party analysis/computation service.
Access to the GRADLC (\url{https://gradlc-dc.iap.kit.edu}) open data accumulator is on the way, and extending the platform by more analysis options and data shops are a worthwhile consideration.
All in all Jupyter proved to be a great tool for education and outreach, at least.

\section{KCDC API}
Starting with the SKARAGAN release KCDC provides users, in addition to a graphical interface, an API (Application Programming Interface) for receiving data.
The API gives users the ability to customise data processing at a higher level. In particular, the user can automate many stages of downloading data and achieve more flexible control over downloading, e.g.~by configuring pending requests, downloading data by condition, downloading data in parts, monitoring the status of dispatched requests, and more.
In the case of relatively small requests, the API can be used directly in the Jupiterhub system when loading data for analysis. The KCDC API supports popular RESTful (REpresentation State Translate) API design. The data is transferred using the HTTPS protocol. The protocol supports actions like: providing information about available data, sending a request, deleting a request, and checking its status.
A large number of data models are supported, such as: Request, Dataset, Quantity, Cut, etc. More information about the API can be found in the KCDC User Manual for the KASCADE-Grande DataShop (\url{https://kcdc.iap.kit.edu/static/pdf/kcdc_mainpage/kcdc-Manual.pdf}) and in the API documentation (\url{https://kcdc.iap.kit.edu/datashop/api/docs/index.html}).

Let us look at a request example using the KCDC API with the Linux command line. For example, a user wants to retrieve all the data within the energy range $10^{17}$--$10^{19}$ eV (17--19 [log10 eV]). Then the following query
\begin{footnotesize}
\begin{verbatim}
curl --insecure --request POST 'https://kcdc.iap.kit.edu/datashop/api/submit' \
--header 'Authorization: Basic cG92dGVyOmhhcnJ5Kytxb3R0ZXI=' \
--header 'Content-Type: application/json' \
--data-raw '
{
    "reconstruction": "",
    "output_format": "ascii",
    "datasets": [
        {
            "name": "array",
            "quantities": [
                {
                    "name": "E",
                    "cuts": [[17, 19]]
                }
            ]
        }
    ]
}'
\end{verbatim}
\end{footnotesize}
where username:password are the user's KCDC credentials in base64 encoding, will return a response with the unique identifier of the accepted request or an error message in case of an invalid request. The same as in case of working with graphical user interface, when the request is completed, a message about and a download link is sent to the email specified by the user during registration.

\section{Education and Training}
Materials from the KCDC portal were used while carrying out events like the International Cosmic Day (ICD) 2018~\cite{icd} and 2020~\cite{kcdc_mk}, as well as for preparing lectures and training materials as part of the collaborative work with colleagues from GRADLC (\url{https://gradlc-dc.iap.kit.edu}), which was reflected in some materials on the astroparticle.online portal (\url{https://astroparticle.online/})~\cite{astrop_online}.
\begin{figure}[htbp]
\begin{minipage}[h]{0.47\linewidth}
\center{\includegraphics[width=1\linewidth]{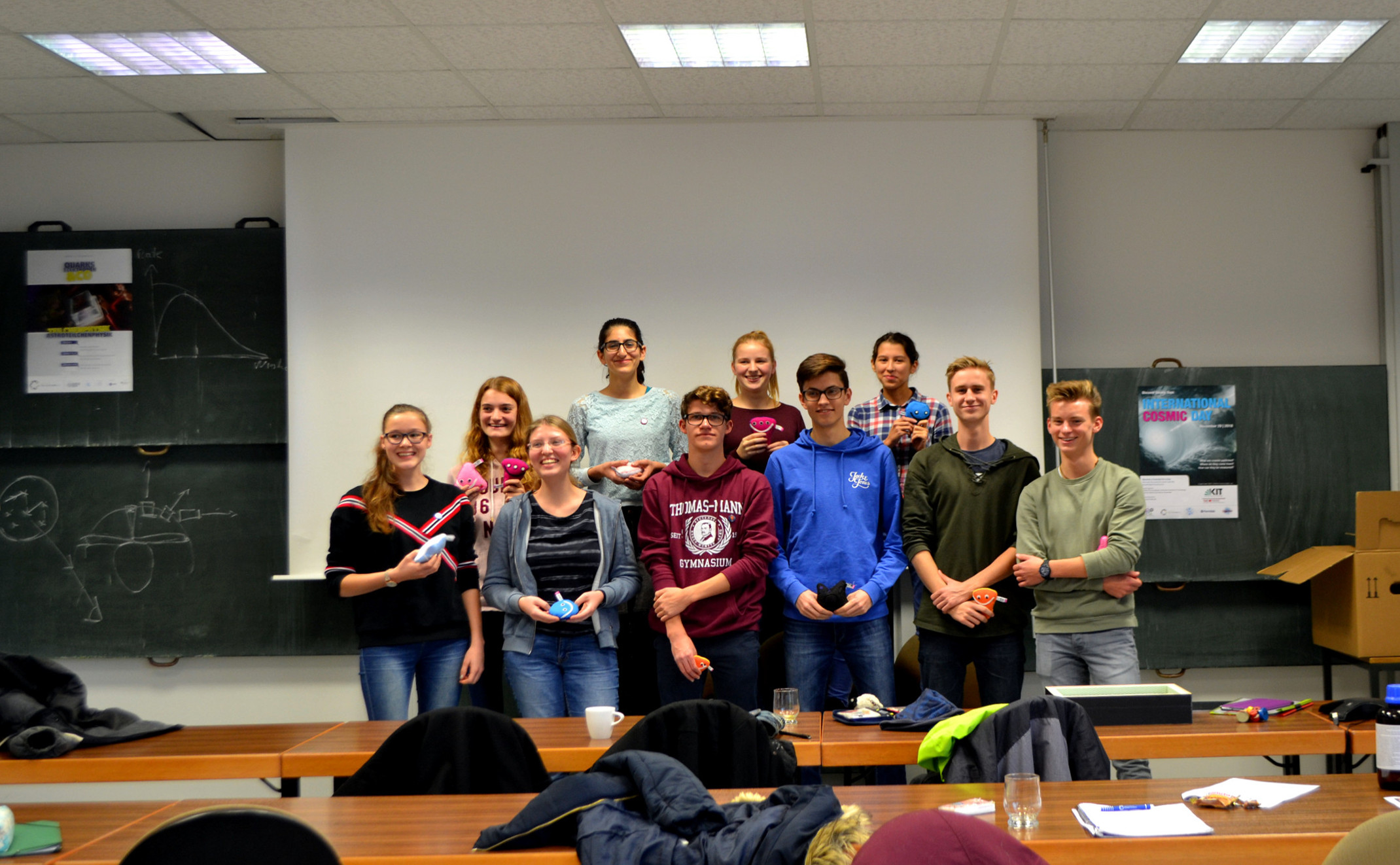}} a) \\
\end{minipage}
\hfill
\begin{minipage}[h]{0.47\linewidth}
\center{\includegraphics[width=1\linewidth]{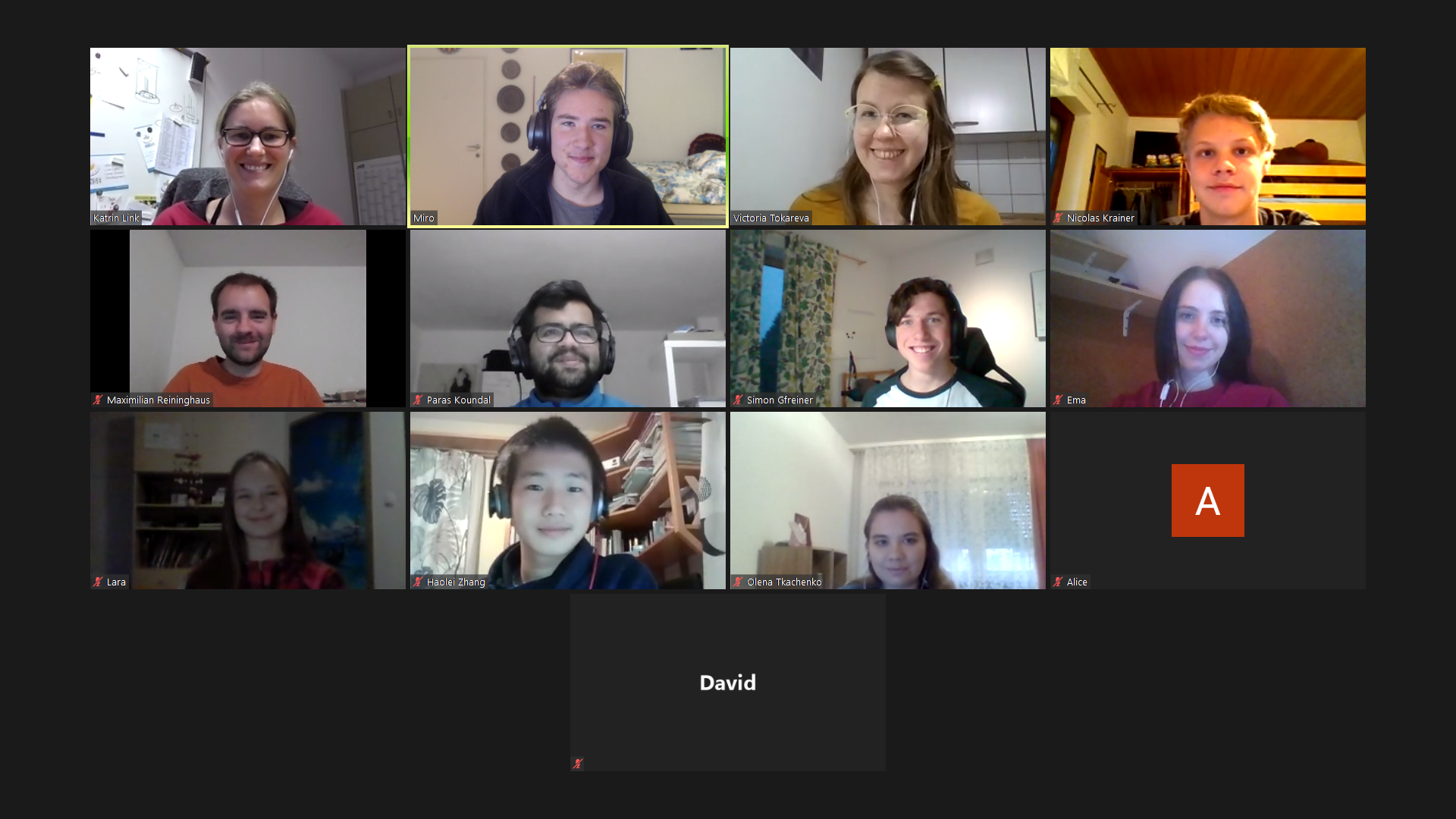}} \\b)
\end{minipage}
\caption{
  a) Group photo of ICD-18 participants;
  b) Group photo of ICD-20 participants;
}
\end{figure}

The use of the portal's resources turned out to be especially intensive during the masterclass for ICD-20, since the masterclass had to be performed completely online due to the pandemic situation. To deal with that, the following actions were performed:
\vspace{-\topsep}
\begin{itemize}
\setlength{\itemsep}{0pt}
\setlength{\parskip}{0pt}
\setlength{\parsep}{0pt}
\item pre-selecting data for analysis
\item preparing interactive notebooks with exercises
\item preparing instructions to work with data and perform exercises
\item giving lectures on the basics of cosmic-ray physics and programming: it was shown online how to work with the KCDC portal and analyse data on the Jupyterhub platform
\end{itemize}
\vspace{-\topsep}
The developed materials were published as publicly available tutorials in KCDC's Jupyterhub environment.

\section{Summary and Outlook}

Emerged from the requests of the community and our experience on how to handle big data in science, we have several tasks and ideas on our to-do list for the further development of KCDC.
In the next release (\textbf{SKARAGAN 2.0}) scheduled for June 2021, a complete upgrade of all software components will be published, based on UBUNTU 20.04 LTS.
Most of all, a replacement for the ftp download server is necessary because ftp is already deprecated in modern browsers and a download of user requests and simulations is not supported any more. 

With the next major release in the pipeline called~\textbf{QUALOR}, we want to completely replace the sharded cluster to speed up the processing of user requests. For this, a complete new filling of the MongoDB will be necessary.

\begin{small}
\paragraph{Acknowledgements:}
Supported by KRAD, the Karlsruhe-Russian Astroparticle Data Life Cycle Initiative (Helmholtz Society Grant  HRSF-0027). The authors acknowledge the cooperation with the Russian colleagues (A.~Kryukov et al.) in the GRADLC project (RSF Grant No. 18-41-06003) as well as the KASCADE-Grande collaboration  for their continuous support of the KCDC project.
We are very grateful to Ashot Chilingarian and Gagik Hovsepyan for the permission to publish the Maket-Ani data in KCDC and for their help in preparing the data sets and the documentation.
\end{small}

\begin{small} 

\end{small} 

%

%
%
%


\begin{thebibliography}{99}

\bibitem{kcdc_paper}
A. Haungs et al; 'The KASCADE Cosmic-ray Data Centre KCDC: Granting Open
Access to Astroparticle Physics Research Data'; Eur. Phys. J. C (2018) \textbf{78}:741;
\url{https://doi.org/10.1140/epjc/s10052-018-6221-2}

\bibitem{kang_kascade_icrc2021} 
KASCADE-Grande Collaboration, D. Kang et al.; Results from the KASCADE-Grande data analysis; PoS ICRC2021 (these proceedings) ID 565

\bibitem{LOPES}
H. Falcke et al; ‘Detection and imaging of atmospheric radio flashes from cosmic ray air showers’; Nature \textbf{435}:313 (2005); \url{https://doi.org/10.1038/nature03614},

\bibitem{lopes-new}
W.D. Apel et al., LOPES Collaboration; 'Final results of the LOPES radio interferometer for cosmic-ray air showers'
Eur.Phys.J.C \textbf{81} 2, 176 (2021)
\url{https://doi.org/10.1140/epjc/s10052-021-08912-4} 

\bibitem{MATOMO}
MATOMO analytics ; \url{https://matomo.org/100-data-ownership/}

\bibitem{apj_603-2004}
A. Chilingarian, G. Gharagyozyan, G. Hovsepyan, S. Ghazaryan, L. Melkumyan, and A. Vardanyan; Light and Heavy Cosmic Ray Mass Group Energy Spectra as Measured by the MAKET-ANI Detector; Astrophysical Journal, \textbf{603}:L29-L32, (2004); \url{https://doi.org/10.1086/383086/}.

\bibitem{ap_28-2007}
A. Chilingarian, G. Gharagyozyan, G. Hovsepyan, S. Ghazaryan, L. Melkumyan, A. Vardanyan, E. Mamidjanyan, V. Romakhin, and S. Sokhoyan; Study of extensive air showers and primary energy spectra by MAKET-ANI detector on Mount Aragats; Astroparticle Physics, \textbf{28}, Issue 1, September 2007, Pages 58–71; \url{https://doi.org/10.1016/j.astropartphys.2007.04.005}.

\bibitem{icd} 
The 9th International Cosmic Day 
website; \url{https://icd.desy.de/}; accessed June 2021.

\bibitem{kcdc_mk} 
K. Link, V. Tokareva, A. Haungs, D. Kang, P. Koundal, F. Polgart, O. Tkachenko, D. Wochele, and J. Wochele; 'Online masterclass built on the KASCADE cosmic ray data centre'; PoS ICRC2021 (these proceedings) 1378.

\bibitem{astrop_online} 
V. Tokareva, Y. Kazarina, V. Samoliga, A. Haungs, A. Kryukov, E. Postnikov, A. Shigarov, D. Shipilov, and D. Zhurov; 'Multi-messenger astroparticle physics for the public via the astroparticle.online project'; PoS ICRC2021 (these proceedings) 1373.

\bibitem{gradlc} 
V. Tokareva et al., 'German-Russian Astroparticle Data Life Cycle Initiative to foster Big Data Infrastructure for Multi-Messenger Astronomy';  PoS ICRC2021 (these proceedings) 938.

\end{thebibliography}
\end{document}